\documentclass[%
 reprint,
 superscriptaddress,
 amsmath,amssymb,
 aps,
 prl,
]{revtex4-2}
\usepackage{xcolor}
\usepackage{lipsum} % for dummy text only
\newcommand{\expe}[1]{\left\langle #1 \right\rangle}

\usepackage{appendix}
\usepackage{braket}
\usepackage{float} 
\usepackage{chemformula}
\usepackage{comment}
\usepackage{graphicx}% Include figure files
\usepackage{dcolumn}% Align table columns on decimal point
\usepackage{bm}% bold math
\usepackage{setspace}
\usepackage{braket}
\usepackage{graphicx}% Include figure files
\usepackage{dcolumn}% Align table columns on decimal point
\usepackage{bm}% bold math
\usepackage[colorlinks=true, citecolor=blue, linkcolor=blue, urlcolor=blue]{hyperref}

\makeatletter
\newcommand*{\addFileDependency}[1]{% argument=file name and extension
  \typeout{(#1)}
  \@addtofilelist{#1}
  \IfFileExists{#1}{}{\typeout{No file #1.}}
}
\makeatother

\setcitestyle{super}

\begin{document}

\preprint{APS/123-QED}
\title{Giant Resonant Enhancement of Photoinduced \texorpdfstring{\\}{ } Dynamical Cooper Pairing,  \textit{far above $T_c$}}

\newcommand{\affiliationHavard}{
Lyman Laboratory, Department of Physics, Harvard University, Cambridge, MA 02138, USA
}

\newcommand{\affiliationMPSD}{
Max Planck Institute for the Structure and Dynamics of Matter,
Center for Free-Electron Laser Science (CFEL),
Luruper Chaussee 149, 22761 Hamburg, Germany
}

\newcommand{\affiliationETH}{
Institute for Theoretical Physics, ETH Z\"urich, 8093 Z\"urich, Switzerland
}

\newcommand{\affiliationOxford}{
Department	of	Physics,	Clarendon	Laboratory,	University	of	Oxford,	United	Kingdom
}

\newcommand{\affiliationPKS}{Max Planck Institute for the Physics of Complex Systems, Nöthnitzer Straße 38, 01187 Dresden, Germany}

\author{Sambuddha Chattopadhyay}
\affiliation{\affiliationHavard}
\affiliation{\affiliationETH}

\author{Marios H. Michael}
\affiliation{\affiliationPKS}
\affiliation{\affiliationMPSD}

\author{Andrea Cavalleri}
\affiliation{\affiliationMPSD}
\affiliation{\affiliationOxford}

\author{Eugene A. Demler}
\affiliation{\affiliationETH}
\date{\today}

\begin{abstract}

Pump-probe experiments performed on $\mathrm{K}_3\mathrm{C}_{60}$ have unveiled both optical and transport signatures of metastable light-induced superconductivity up to room temperature, far above $T_c$. Recent experiments have uncovered that excitation in the vicinity of $50 \textrm{meV}$ enables the observation of high temperature light-induced superconductivity at significantly lower fluences. Inspired by these experiments we develop a mechanism which can explain such a giant resonant enhancement of light-induced superconductivity. Within a minimal non-linear Holstein model, we show that resonantly driving optical Raman modes leads to a time-dependent electron-phonon coupling. Such a coupling then modulates the effective electron-electron attraction, with the strongest modulations occurring when the drive is resonant with the phonon frequency. These dynamical modulations of the pairing interactions lead to Floquet-BCS instabilities at temperatures far exceeding equilibrium $T_c$, as observed in experiments. We conclude by discussing the implications of our general analysis on the $\mathrm{K}_3\mathrm{C}_{60}$ experiments specifically and suggesting experimental signatures of our mechanism.
\end{abstract}
\maketitle

\textbf{\textit{Introduction.}}---The optical control of emergent properties in quantum materials is a focal project of contemporary solid state physics \cite{basov_towards_2017,de_la_torre_colloquium_2021}. Among the most striking developments is the photocontrol of superconductivity, highlighted by optical \cite{Mitrano_16} and terahertz transport \cite{joe_i,joe_ii} evidence for nanosecond-scale metastable \cite{Budden_21} light-induced superconductivity in alkali-doped fulleride \ch{K3C60} observed up to room temperature, far above the equilibrium $T_c=19{\rm K}$ \cite{Gunnarsson_04}. The experiments have sparked intense theoretical activity on both the microscopic origin of high-temperature pairing \cite{kennes_transient_2017,komnik_bcs_2016,nava_cooling_2018,knap_dynamical_2016,kim_enhancing_2016,murakami_nonequilibrium_2017,sentef_theory_2016,babadi_theory_2017,dasari_transient_2018,mazza_nonequilibrium_2017,biswa_i} and its metastability \cite{biswa_i,grankin}: Both aspects remain unresolved. Crucially, the \ch{K3C60} experiments pose the challenge of explaining how photoinduced pairing develops at temperatures fifteen times higher than the equilibrium $T_c$ in \ch{K3C60} which is set by the disappearance of local order\cite{nernst}.  

Recent work (Ref.~\citenum{Ed_23}) has revealed a giant resonant enhancement of light-induced superconductivity: driving \ch{K3C60} near $50$ ${\rm meV}$ produces optical signatures of pairing at fluences two orders of magnitude smaller than earlier experiments at $170$ ${\rm meV}$ \cite{Mitrano_16,Budden_21}. This was quantified by a photo-susceptibility—defined as the normalized inverse fluence needed to elicit a superconducting optical response—which reaches unity when superconductivity is observed at minimal fluence. These resonances provide the sharpest microscopic clues into the origin of high-temperature photoinduced pairing in \ch{K3C60}.

However, the observed resonances pose interpretive challenges: they are broad ($24$–$86\mathrm{meV}$), relatively flat around $40$–$60\mathrm{meV}$, and not attributable to a single IR phonon which have linewidths $\sim 1 {\rm meV}$. Instead, their breadth coincides well with the Raman spectrum of \ch{K3C60}, dominated by $\mathrm{Hg}$ phonons spanning $18$–$100\mathrm{meV}$ \cite{raman,Gunnarsson_04}. We build a microscopic mechanism from this observation in which parametric driving of the Raman ${\rm Hg}$ modes dynamically modulates the pairing interaction. This modulation triggers Floquet-BCS instabilities \cite{knap_dynamical_2016,babadi_theory_2017} enabling Cooper pairing at $T_c^\star \gg T_c$. Our work qualitatively reproduces the giant, broad photo-susceptibility features observed in experiments and provides a generic route to photoinduced superconductivity in driven electron-phonon systems.

\textbf{\textit{Model.}}—We consider a minimal setting of electrons locally coupled to a single branch of dispersionless Raman phonons, capturing essential ingredients of alkali-doped fullerides: narrow bands and strong local electron–phonon interactions with high-frequency intramolecular modes \cite{Gunnarsson_04}. Our Hamiltonian reads
\begin{equation}
H = H_{\rm el} +  H_{\rm el-ph}+ H_{\rm ph}(t),
\label{eq:model}
\end{equation}
where $H_{\rm el}$ describes nearest-neighbor hopping $-t\sum_{\langle ij\rangle,\sigma} (c^\dagger_{i\sigma} c_{j\sigma} + h.c.) - \mu\sum_i n_i$ and $H_{\rm el-ph} = \sum_i g (1+\epsilon Q_i\ell_0^{-1}) Q_i n_i$ encodes a weakly \textit{non-linear Holstein} coupling\cite{nonlin-e-ph}. As Raman modes are inversion-symmetric, linear driving of the phonons is forbidden. From symmetry considerations we write down the Hamiltonian $ H_{\rm ph}(t)=\sum_i \Big(\frac{P_i^2}{2M}+\frac{M\omega^2}{2}Q_i^2+\frac{\lambda}{12\ell_0^4}Q_i^4+\beta E^2(t)Q_i^2\Big)$, where $E(t)=E_0\cos(\omega_{\rm dr}t)$. The parametric coupling $\beta E^2 Q^2$ is resonant when $\omega_{\rm dr}\!\approx\!\omega$ and for sufficiently large $E_0$, the phonon develops large coherent oscillations $\langle Q(t) \rangle=Q_{\rm ss}\cos(\omega_{\rm dr}t)$, stabilized by anharmonicity (see Appendix A). We choose $\beta>0$, the non-trivial case. A semi-classical analysis gives the steady-state amplitude $Q_{\rm ss} = \sqrt{\frac{1}{\lambda}\big(\sqrt{\kappa^2-\gamma^2} -\Delta \big)}\ell_0$\cite{milburn}; where $\kappa = \beta E_0^2/M\omega_{\rm eff}$; where the detuning $\Delta = \omega_{\rm eff}-\omega_{\rm dr}$; where $\omega_{\rm eff} = \sqrt{\omega^2+\beta E_0^2/M}$, capturing the shift in the phonon frequency arising from the DC component of the parametric drive.

By linearizing the fluctuations of the phonon around the driven steady state given by $\langle Q(t)\rangle$, the effective electron–phonon coupling reduces to \textit{a time-dependent Holstein model},
\begin{equation}
H_{\rm el-ph}(t) = g(1+A_0\cos(\omega_{\rm dr}t))\sum_i \tilde Q_i n_i,
\label{eq:holstein-td}
\end{equation}
where $A_0 =  2\epsilon Q_{\rm ss}\ell_0^{-1}$ sets the modulation strength. To leading order in the anti-adiabatic limit ($\omega \gg t$) relevant to \ch{K3C60}, a time-dependent Schrieffer–Wolff transformation yields an effective pairing interaction
\begin{equation}
U(t) = \frac{g^2}{2M\omega^2}(1+A_0\cos(\omega_{\rm dr}t))(1+A_0\cos(\omega_{\rm dr}t)\mathcal{R}(\omega_{\rm dr})),
\label{eq:Utime}
\end{equation}
with $\mathcal{R}(\omega_{\rm dr})=\omega^2/((\omega^2-\omega_{\rm dr}^2)^2+\gamma^2\omega_{\rm dr}^2)^{1/2}$, where $\gamma$ is the linewidth of the phonon (see Appendix B for details). Near resonance $\omega_{\rm dr}\approx\omega$, $\mathcal{R}$ is sharply enhanced  $\sim \mathcal{Q}\equiv\omega/\gamma$, giving rise to giant modulations of the attractive interaction. These modulations trigger Floquet–BCS instabilities at electronic temperatures $T_c^\star$ far above the equilibrium $T_c$ \cite{knap_dynamical_2016,babadi_theory_2017}. A pedagogical derivation from the microscopic model in Eq. ~\ref{eq:model} to Eq.~\ref{eq:Utime} is presented in the Appendix.

\begin{figure}[t!]
\includegraphics[width=\columnwidth]{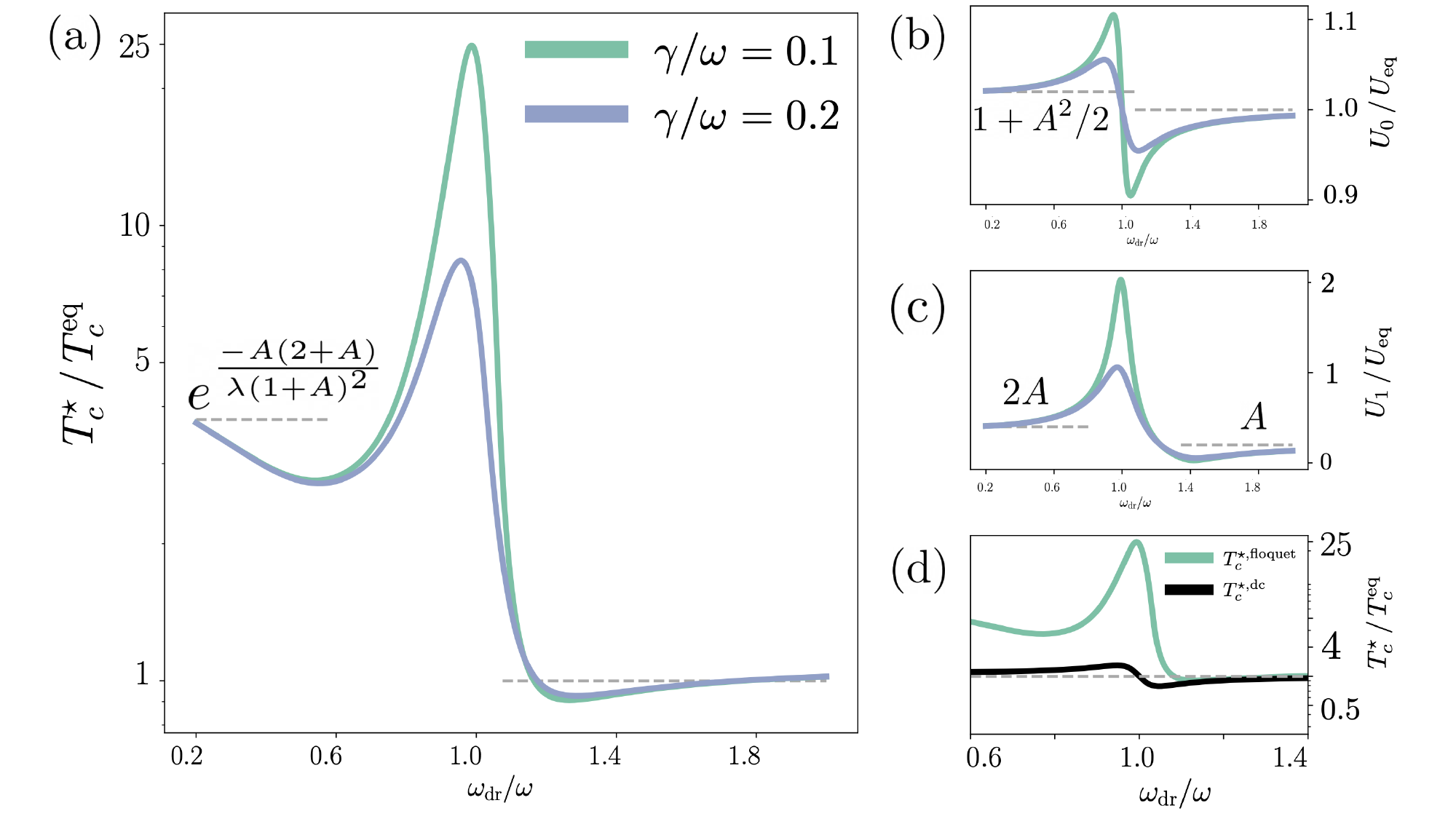}
\caption{\textbf{High-$T_c^\star$ Dynamical Cooper Pairing in the Time-Dependent Holstein Model.} (a) $T_c^\star/T_c$---obtained by solving  Eq.~\ref{eq:floquet_bcs_equation} within the  the time-dependent Holstein model (Eq.~\ref{eq:time_dependent_holstein}) with $A_0 = 0.2$---for two different $\mathcal{Q} = \frac{\omega}{\gamma}$ factor phonon modes. Dashed gray lines refer to analytic asymptotics at low (see text) and high frequencies (convergence to $T_c$). (b) The static contribution to the interaction $U_0/U_{\rm eq}$---where $U_{\rm eq} = U_{\rm bi}$---showing (small) resonant enhancement (suppression) under red-detuned (blue-detuned) driving. (c) The AC contribution to the interaction at the drive frequency $U_1/U_{\rm eq}$ showcasing \textit{giant}, unsigned enhancement in the vicinity of resonant driving. This resonant contribution \textit{drives} the resonant enhancement of $T_c^\star$. (d) A comparison---for $\mathcal{Q} = 10$---between $T_c^\star$ from a full Floquet calculation (teal) and $T_c(U_0)$ obtained within BCS theory (black), underscoring the Floquet nature of the BCS instability.}
\label{fig:fig1}
\end{figure}

\textbf{\textit{Dynamical Cooper Pairing \& the Floquet-BCS Instability.}}---Instability drives the formation of order in classical and quantum systems\cite{Cross_Greenside_2009, AGD}. For orientation, it is useful to recall the situation for superconductivity in equilibrium within a simplified Landau-Ginzburg picture where at $T_c$, the free energy landscape flips in concavity in the vicinity of $\Delta = \frac{1}{N}\sum_k \langle c^\dag_{k \uparrow} c^\dag_ {-k \downarrow} \rangle=0$. Dynamically, this causes the order parameter to grow exponentially until non-linearities saturate its formation\cite{AGD}. In what follows, we calculate the temperature $T_{c}^\star$---a non-equilibrium analogue of equilibrium $T_c$ in the sense outlined above---below which the Floquet-BCS instability takes place. We use the formalism pedagogically articulated in Refs \citenum{knap_dynamical_2016, babadi_theory_2017} which leverages a Floquet \textit{ansatz} within a time-dependent Hartree-Fock approximation to describe the initial evolution of $\Delta$, deriving a set of self-consistency equations for the instability rate $\Gamma_{\rm pair}$. We solve these equations numerically around $\Gamma_{\rm pair} = 0$ to extract $T_c^\star$.

%Instability drives the formation of order in classical and quantum systems\cite{Cross_Greenside_2009, AGD}. For orientation, it is useful to recall the situation for superconductivity in equilibrium within a simplified Landau-Ginzburg picture where at $T_c$, the free energy landscape flips in concavity in the vicinity of $\Delta = \frac{1}{N}\sum_k \langle c^\dag_{k \uparrow} c^\dag_ {-k \downarrow} \rangle=0$. Dynamically, this causes the order parameter to grow exponentially until non-linearities saturate its formation\cite{AGD}. In what follows,

We now specify the calculation outlined above. Having integrated out the fluctuations of the high-frequency, driven phonons , our electronic model is given by: $H(t) = H_{\rm el}- U(t)\sum_{i} n_i^2$ \footnote{We neglect Hartree contributions to the chemical potential as they can be gauged away. \cite{knap_dynamical_2016}}. We are interested in the dynamics of $\Delta(t) = \frac{1}{N}\sum_k \langle c^\dag_{k \uparrow} c^\dag_ {-k \downarrow} (t)\rangle \approx e^{\Gamma_{\rm pair}t}\sum_n \Delta_n \exp(i n \omega_{\rm dr}t)$, within a Floquet ansatz. At $T_c^\star$---where $\Gamma_{\rm pair}$ becomes finite---the Floquet-BCS instability equation is given by the following infinite system of equations: 
\begin{equation}
(1+U_0 F_n) \Delta_n - \frac{U_1}{2} F_n (\Delta_{n-1}
+ \Delta_{n+1})\\=0    
\label{eq:floquet_bcs_equation}
\end{equation}
where $U(t) = U_0+U_1 \cos(\omega_{dr} t)$; where $F_n = \lim _{\eta \to 0} \nu(0) \int_{-W/2}^{W/2} d\xi \frac{\tanh{(\xi/(2 T)})}{2\xi - n\omega_{\rm dr}+ i\eta}$, where $\nu(0)$ is the density of states at the Fermi energy and $W$, the electronic bandwidth, sets the UV cutoff\cite{micnas} and $U_0$ and $U_1$ are obtained from Eq.~\ref{eff_attraction}. We neglect AC contributions to $U(t)$ at $2\omega_{\rm dr}$ as they are higher order in $A_0 \ll 1$. 

\begin{comment}
    \begin{widetext}
\begin{align}
U_0 &= 
 \frac{g^2}{2 M \omega^2}\Big(1 + \frac{A^2}{2}
    \frac{\omega^2 (\omega^2 - \omega_{\rm dr}^2)}{(\omega^2 - \omega_{\rm dr}^2)^2 + (\gamma\omega_{\rm dr})^2}\Big)\\
U_1 &= \frac{g^2 A}{2 M \omega^2}\sqrt{\Big(1+\frac{\omega^2 \big(\omega^2-\omega_{\rm dr}^2\big)}{\big(\omega^2-\omega_{\rm dr}^2\big)^2+\big(\gamma \omega_{\rm dr}\big)^2}\Big)^2+\Big(\frac{\gamma \omega_{\rm dr}\omega^2}{\big(\omega^2-\omega_{\rm dr}^2\big)^2+\big(\gamma \omega_{\rm dr}\big)^2}\Big)^2}
\label{eq:U0_U1}
\end{align}
\end{widetext}
\end{comment}

\begin{figure}[t!]
\includegraphics[width=\columnwidth]{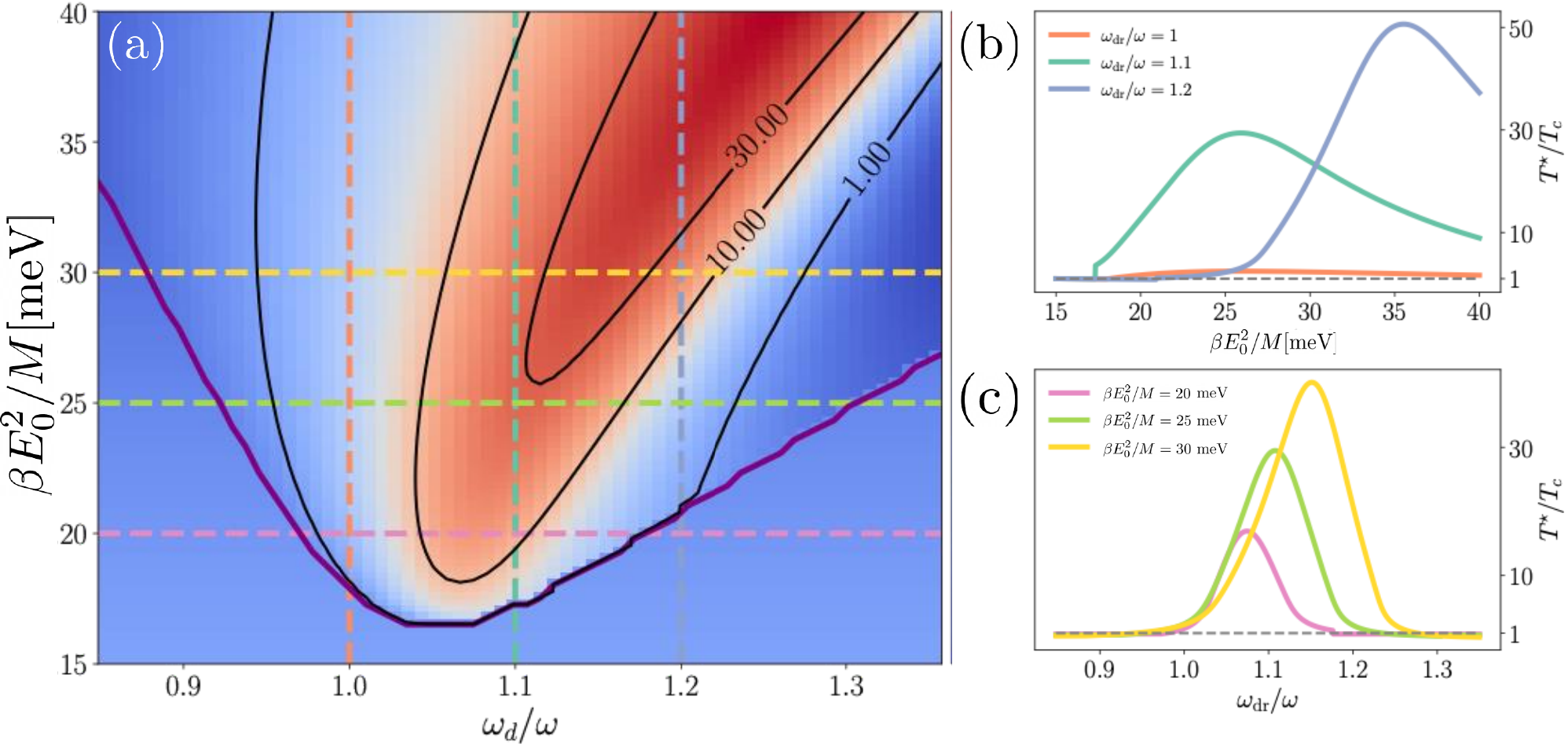}
\caption{\textbf{Giant Resonant Enhancement of Photoinduced Cooper Pairing \textit{via} Parametric Driving} (a) $T_c^\star/T_c$ shown as a function of frequency $\omega_{\rm dr}/\omega$ and the energy scale associated with the electric field $\beta E_0^2/M$. Black lines reveal contours ($1\times, 10 \times, 30 \times$ enhancement); the purple line demarcates the parametrically unstable region; orange, green, blue are the cuts used to generate (b); yellow, green, pink dashed lines are the cuts used to generate (c). (b) $T_c^\star/T_c$ at fixed $\omega_{\rm dr}/\omega$ as a function of the electric field. (c) $T_c^\star/T_c$ at fixed $E_0$, as a function of normalized drive frequency.}
\label{fig:fig2}
\end{figure}

In Fig.~\ref{fig:fig1}, we provide an illustrative set of examples of the giant resonant enhancement of $T_{c}^\star$ for fixed $A_0$. In Fig.~\ref{fig:fig1}a, we plot  $T_c^\star$ as a function of $\omega_{\rm dr}/\omega$ for $A_0=0.2$, corresponding to plausible estimates for $\epsilon$ and $Q_{\rm ss}$ (e.g., $\epsilon = 0.05$, $Q_{\rm ss}\ell_0^{-1} = 2$). In Fig.~\ref{fig:fig2}, we present our results within the broader microscopic context (i.e., incorporating the effects of parametric driving). In both plots we choose parameters that economically characterize equilibrium superconductivity in \ch{K3C60} within a BCS framework: $T_c = 19$K, $\omega = 50 {\rm meV}$,  $W = 500{\rm meV}$, and $\lambda_{\rm BCS} = \nu(0)U_{\rm bi} = 0.2$. For Fig.~\ref{fig:fig2}, we choose  $\lambda = 0.1 \omega$ for the phonon anharmonicity and $\epsilon = 0.05$ for the non-linear contribution to the electron-phonon coupling. 

Strikingly, in Fig.~\ref{fig:fig1}a, we show that on resonance, for $\mathcal{Q}=20$, $T_c^\star$ is over $20$ times larger than $T_c$, as is consistent with experiments. Similar to what previous theoretical studies in the driven quantum materials context \cite{denis, dieter, sunny} have underscored, under driving the static part of the interaction $U_0$ is resonantly modified with red-detuned (blue-detuned) driving increasing (decreasing) the effective bipolaronic energy\cite{denis, sunny}. As shown in Fig.~\ref{fig:fig1}(b), this contribution ( $T_c^{\rm dc}$) is signed but, as plotted in black Fig.~\ref{fig:fig1}(d), does not give rise to the resonant enhancement of $T_c^\star$ (teal). What underlies the resonant enhancement of $T_c^\star$ instead is the unsigned, resonant enhancement of $U_1$ (depicted in Fig.~\ref{fig:fig1}(c)), the AC modulation of the interaction. Formally, the dominance of the resonant modification of $U_1$ over that of $U_0$ arises from the fact that contributions to $U_0$ are second order in perturbation theory in $A_0$, while contributions to $U_1$ are linear in $A_0$. Thus in assessing the instability temperature it is \textit{crucial} to account for the Floquet nature of the instability. We emphasize that the frequency dependence arising from the Floquet nature of the dynamical Cooper pairing instability is consistent with experiments which do \textit{not} find large dips in photo-susceptibility in blue-detuned regions \cite{Ed_23}, as driven modifications of $U_0$ alone would suggest. We note in passing that the increase of $T_c^\star$ at low frequencies can be understood as the BCS critical temperature arising from $U_{\rm eff}^{\rm dc} = U_0+U_1$. While it is interesting that a DC heuristic yields a semi-quantitative estimate for $T_c^\star$ at frequencies $\omega_{\rm dr} \sim 0.2 \omega$ for fixed $A_0$ we do not recover this low-frequency physics within the calculations displayed in Fig.~\ref{fig:fig2} as this behavior occurs at frequencies that are outside of the parametrically unstable regime. 

In Fig.~\ref{fig:fig2}a, we compute $T_c^\star$ as a function of the electric field $E_0$ and driving frequency $\omega_{\rm dr}$ in the vicinity of $\omega$ , considering the underlying parametric instability that sources the coherent oscillations of the phonons, computing $A_0(E_0, \omega_{\rm dr})$.  We capture the interplay between Floquet enhancement of $T_c^\star$ and the lowering of the bipolaron energy $U_{\rm bi}$---which impacts both static and modulated components of $U(t)$---arising from phonon hardening due to the DC component of the electric field $E_0$ ($\omega_{\rm eff}(E_0)$). The discrepancy between the parametrically unstable line (purple) and the break-even line (black)---where the driven instability temperatures matches the undriven equilibrium critical temperature $T_c^\star = T_c$---highlights that even when the phonon is parametrically unstable and there are Floquet contributions to pairing, it does not guarantee that the critical temperature for the pairing instability will increase. A prominent feature is the blue shift of the eye of the plot, an effect that arises from phonon hardening. Note that even within the constraints placed on the phenomenology of the time-dependent Holstein model, the 15 fold increase of $T_c^\star$ observed in experiments is attainable within our calculations in the vicinity of the phonon resonance.

\begin{figure}[t!]
\includegraphics[width=\columnwidth]{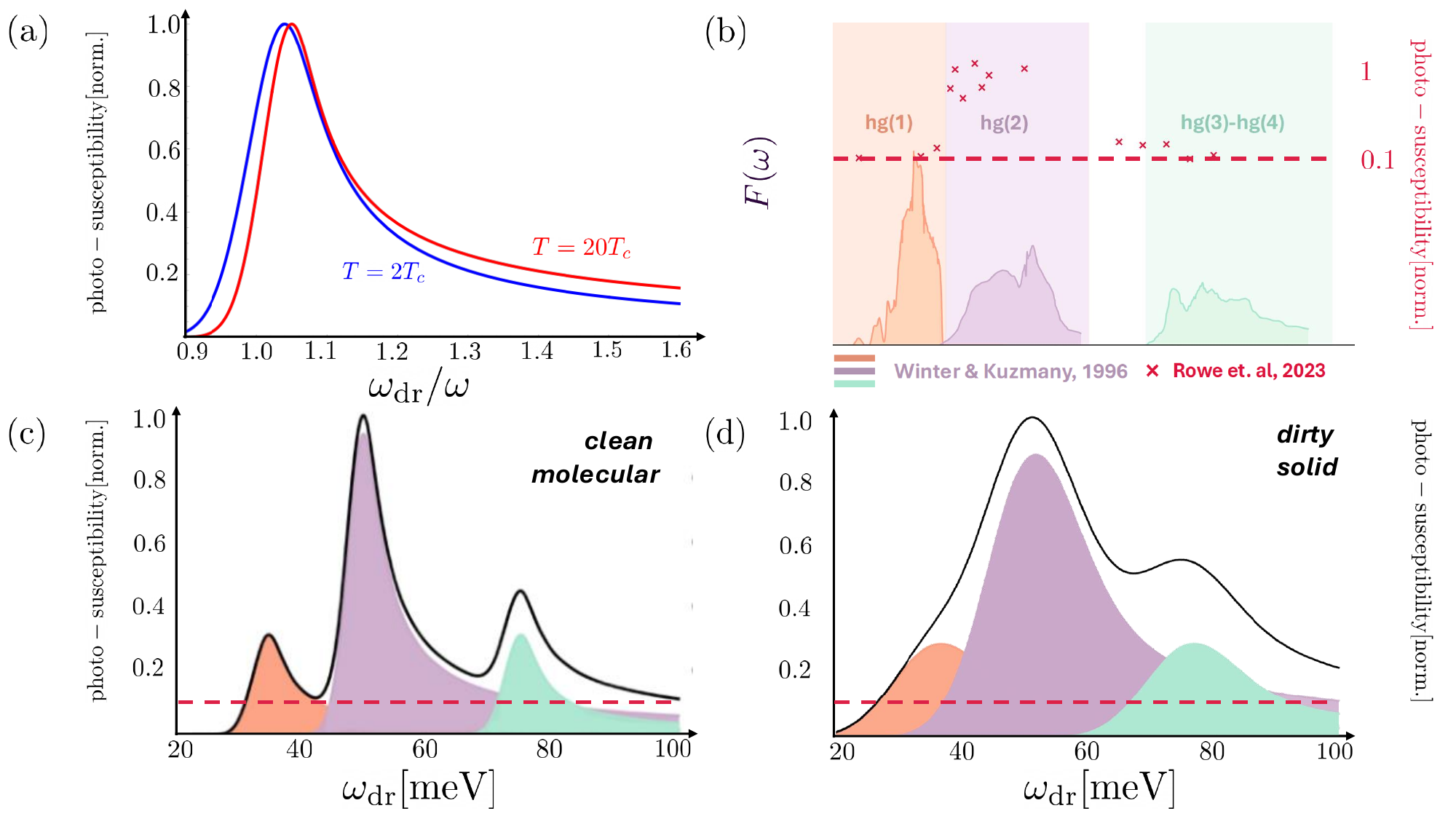}
\caption{\textbf{Photo-susceptibility.} (a) Photo-susceptibility of light-induced superconductivity for the model given by Eq.  \ref{eq:BasicModelFull} for two different temperatures $T = 2T_c$  (blue) and $T=20 T_c$ (red). (b) Schematic of the phononic density of states $F(\omega)$ from historical Raman measurements\cite{raman} in \ch{K3C60} showing broad features which derive from molecular  phonons  ${\rm Hg}$ $1$ (orange), ${\rm Hg}$ $2$ (purple), and ${\rm Hg}$ $3$ and $4$ (light green). Overlain, the raw photo-susceptibility data (crimson crosses) from recent light-induced superconductivity experiments\cite{Ed_23}, largely exhibiting qualitative agreement with the Raman features. (c) Schematic of the photo-susceptibility features arising from individual narrow vibration contributions in the absence of broadening. Contributions from the individual modes are shown in ${\rm Hg}$  $1$ (orange), ${\rm Hg}$ $2$ (purple), and ${\rm Hg}$ $3$ and $4$ (light green). (d) Schematic of the photosusceptibility features arising from broadened vibrational modes showcase a breadth qualitatively consistent with what is measured in experiments.}
\label{fig:fig3}
\end{figure}

\textbf{\textit{Implications for \ch{K3C60}.}}—Having discussed $T_c^\star$, we now turn towards the photosusceptibility features that motivated this study. The broad superconducting photo-susceptibility of \ch{K3C60} coincides closely with its Raman spectrum. 
In Fig.~\ref{fig:fig3}a, using the single-mode model of Eq.~\ref{eq:model} with parameters from Fig.~\ref{fig:fig2}, we compute the photo-susceptibility, defined as  $\frac{E_0^{\star}(\omega_{\rm dr},T)^2}{\max_{\omega_{\rm dr}}(E_0^{\star}(\omega_{\rm dr},T))^2}$ where $E_0^\star$ is the critical electric field required to raise the pairing scale to $T_c^\star=T$. 
For $T/T_c=2$ and $20$ (blue and red curves), the results display three robust features: a sharp resonant peak, a long off-resonant tail from strong-field parametric driving, and a striking insensitivity to temperature apart from a small phonon-hardening shift. 

In the experimental data of Ref.~\citenum{Ed_23}, the resonances appear not as one broad $60$~meV-wide feature but as three distinct clusters centered at $24$–$38$, $41$–$53$, and $70$–$86$~meV (Fig.~\ref{fig:fig3}b). 
This structure matches well with the ${\rm Hg}$ phonons observed in Raman scattering—${\rm Hg}_1$ (18–35~meV), ${\rm Hg}_2$ (37–56~meV), and ${\rm Hg}_{3,4}$ (81–100~meV)—which are widely believed to be the dominant pairing modes in equilibrium \ch{K3C60} \cite{Gunnarsson_04}. 
The observed linewidths of $10$–$15$~meV are much larger than the $\sim 2$~meV pulse bandwidth, but are consistent with broadening seen in Raman spectra, arising from crystal-field splitting of the fivefold ${\rm Hg}$ polarizations, strong electron–phonon coupling, and disorder. 
The homogeneous linewidths remain narrower than the pulse bandwidth, suggesting that our theoretical analysis in the high-$\mathcal{Q}$ limit is an appropriate description of the underlying physics. 

To address the multi-mode character of \ch{K3C60}, Figs.~\ref{fig:fig3}c and \ref{fig:fig3}d schematically extend our single-mode results using \textit{ab initio} estimates of the bipolaron contributions \cite{Gunnarsson_04}. 
In the absence of broadening, separate ${\rm Hg}$ modes would yield distinct sharp resonances, as illustrated in Fig.~\ref{fig:fig3}c. 
In practice, however, the heavy-tailed nature of the individual responses (Fig.~\ref{fig:fig3}a) and the broadening mechanisms discussed above combine to wash out this substructure, merging the features into a single broad susceptibility peak, as shown in Fig.~\ref{fig:fig3}d. 
This picture provides a natural explanation for the experimental observation of a broad response that nonetheless retains clear clustering tied to the underlying Raman-active ${\rm Hg}$ modes. 

\textbf{\textit{Conclusion.}}—
We have introduced a microscopic mechanism for light-induced superconductivity in \ch{K3C60}, starting from the experimental observation that the photo-susceptibility resonances align with Raman-active ${\rm Hg}$ phonons. 
By analyzing a time-dependent Holstein model generated through parametric driving, we showed that the pairing interaction acquires a resonantly enhanced modulation which triggers Floquet-BCS instabilities at $T_c^\star \!\gg\! T_c$. 
This framework both reproduces the qualitative features of the observed resonances and establishes the time-dependent Holstein model as a paradigmatic description of non-equilibrium phonon-mediated pairing. 

Our proposal can be falsified in several direct ways. Time-resolved Raman or ultrafast electron diffraction can reveal the coherent phonon oscillations predicted here to be $3-5 \ell_0 \sim 10 {\rm pm}$, while systematic frequency- and temperature-dependent measurements of photo-susceptibility can test our predictions of nearly temperature-independent normalized line-shapes and redshifts due to phonon hardening, revealing sharper individual peaks in cleaner samples. These signatures, together with the clustering of resonance energies around the ${\rm Hg}$ modes, provide a concrete set of benchmarks for experiment. 

Looking forward, improved \textit{ab initio} calculations of non-linear couplings ($\epsilon$, $\beta$, $\lambda$) and their interplay across multiple ${\rm Hg}$ modes will be crucial to quantify the feasibility of our mechanism in \ch{K3C60}. There are several theoretical questions opened by our analysis. While in this work we have described the conditions under which the Floquet-BCS instability occurs (i.e., performed a linear stability analysis around the metallic state), it is of interest to track the evolution of the (local) order parameter as non-linearities saturate the growth of this linear instability, asking questions such as ``Do non-linearities suppress the Floquet, oscillating nature of the gap?", ``What is the magnitude---and perhaps temporal variance---of the saturated gap?", ``How does the concomitant evolution of the non-equilibrium distribution of  quasi-particles modify the non-linear dynamics of the Floquet instability?" If oscillations of the gap persist, these oscillations would lead to a time-varying permittivity of the system. It is then of significant experimental consequence to ask what the optical response of such a system may be, an issue partially entertained in Ref.~\citenum{Buzzi_20}. It is also of interest to ask how macroscopic phase coherence can develop under strong drive\cite{k3c60Meissner}. Addressing these issues will clarify how far superconductivity can be pushed beyond equilibrium limits and guide the search for new materials where light can induce aspects of superconductivity at unprecedented temperatures. 

\section{Acknowledgements}
We acknowledge stimulating discussions with M. Buzzi, E. Rowe, Y. Zhu, G. Jotzu, S. Roy, D. B. Shin, M. Mitrano, A. Rubio, \& the dearly departed Assa Auerbach. S.C. is grateful for support from the NSF under Grant No. DGE-1845298 \& for the hospitality of the Max Planck Institute for the Structure and Dynamics of Matter. M.H.M. would like to acknowledge the support from the Alexander von Humboldt Foundation.

\bibliography{references}% 

\onecolumngrid
\appendix
\newpage
\begin{center}
	\textbf{\Large Supplementary Materials}
\end{center}
\normalsize
\setcounter{equation}{0}

In this Supplementary Materials, we have Appendix A and Appendix B. In Appendix A we begin with a minimal microscopic model, derive the time-dependent Holstein model as an effective model, and then use time-dependent Schrieffer-Wolff---in the absence of a bath---to derive the effective time-dependent pairing interaction. In Appendix B, we derive the ``dissipative Schrieffer-Wolff" transformation used to treat the local driven electron-phonon problem in the presence of a dissipative phononic bath. Our ultimate aim is to integrate out the driven phonon fluctuations in the presence of the dissipative bath to obtain a local, interacting electronic model.\\

\subsection{Appendix A: Deriving the Time Dependent Holstein Model}
We begin by specifying a minimal model for electrons coupled to a \textit{single} branch of dispersionless, optical Raman phonons inspired by certain aspects believed\cite{Gunnarsson_04} to be relevant to superconductivity in alkali-doped fullerides and other organic superconductors: narrow electronic bands and strong, local electron-phonon coupling to the high-frequency molecular phonons. Our Hamiltonian $H$ is given by 
\begin{equation}
    H = H_{\textrm{ph}}(t) + H_{\textrm{el}} + H_{\textrm{e-ph}},
\label{eq:BasicModelFull}
\end{equation}
where  the weak nearest neighbor hopping on a three-dimensional lattice is given by the electronic Hamiltonian $H_{\textrm{el}}=-t \sum_{\langle i, j \rangle, \sigma} \big(c_{i, \sigma}^\dag c_{j, \sigma}+ h.c.\big)-\mu \sum_{i} n_i$. Here $c^\dag_{i, \sigma}$ creates an electron with spin $\sigma \in \{\uparrow, \downarrow\}$ at site $i$; we concern ourselves with a partially filled band, characterized by a chemical potential $\mu$. 

As inversion-symmetric Raman modes are forbidden to carry a dipole moment, 
the driven phonon Hamiltonian is restricted by symmetry considerations to: 
\begin{equation}
H_{\textrm{ph}}(t) = \sum_{i} \frac{P_{i}^2}{2 M }+ \frac{M \omega^2}{2} Q_{i}^2+ \frac{\lambda}{12} \ell_{0}^{-4} Q_{i}^4 \\
+\alpha E^2(t) Q_{i} +  \beta  E^2(t) Q_{i}^2
\label{eq:phonons}
\end{equation}
where $P_i$ and $Q_i$ are the canonical momentum and position coordinates for a local (molecular) phonon of frequency $\omega$ and mass $M$ at site $i$; where the phonon oscillator length is given by $\ell_0^{-2} = M\omega$; where $\lambda$ is the energy scale corresponding to the strength of a weak phonon self-anharmonicity; where $\alpha$ minimally characterizes an isotropic Raman tensor; where $\beta$ quantifies non-linear in $Q$ contributions to the linear susceptibility which lead to the  parametric driving of phonons, arising microscopically from, for example, strong electron-phonon coupling or phononic four wave mixing with sum/difference frequency resonant IR phonons. For concreteness, we take $E(t) = E_0 \cos{(\omega_{\rm dr} t)}$. 

Raman processes ($E^2 Q$) are resonant for driving at $\omega_{\rm dr} = \omega/2$ while parametric processes ($E^2 Q^2$) are resonant for $\omega_{\rm dr} = \omega$. In the presence of a stabilizing non-linearity (e.g. due to $\lambda$), the phenomenology arising from these two terms individually---near their respective resonances and at sufficiently strong fields---are not qualitatively different, up to an inconsequential displacive shift. For asymptotically large fields, however, the steady state amplitude $Q_{\rm ss} \sim E$ for parametric driving while $Q_{\rm ss} \sim E^{2/3}$ for Raman,  suggesting that the former may be more  relevant for the strong fields realized in experiments\cite{milburn, dykman}. Moreover, as parametric processes involving Raman modes align more naturally with the observed photo-susceptibility resonances, we focus on them and consider $\omega_{\rm dr}$ in the vicinity of $\omega$.  

%^their relative scale is quantitatively determinable from a combination of $\frac{\alpha \ell_0}{\beta}$  and $E_0$.  
%The empirical fit between the Raman and photo-susceptibility data, however, suggests that, at least for the first two resonant clusters, the second process is dominant---for clarity, we thus focus on parametric driving.Despite this, we note in passing, that Raman processes involving the ${\rm Hg}$ $2$ mode fit the third photo-susceptibility resonance particularly well. 

Formally, we work in the unusual \textit{anti}-adiabatic limit of the electron-phonon system where $\omega > t$. While not precisely quantitative, this limit has furnished significant insight into the counterintuitive, beyond Migdal approximation aspects of (equilibrium) superconductivity in \ch{K3C60} \cite{Gunnarsson_04, assa}. In our analysis, it also provides a crucial scale separation that will facilitate our consideration of the coupled driven electron-phonon system. Our program is accordingly an \textit{anti}-Born-Oppenheimer approach: We will begin by first elucidating the steady state of the fast driven phonon dynamics with \textit{frozen} instantaneous electronic variables, then integrate out the quantum fluctuations of the phonon around its driven steady state to obtain a dynamically modulated, attractive electron-electron interaction. Over the course of this program, the non-linear Holstein model describing the interaction between the electrons and the phonons will be transformed into a \textit{time-dependent} \textit{linear} Holstein model describing the interaction between electrons and phonon \textit{fluctuations} around the driven steady state. 

Following the program outlined above, we begin by understanding the steady state of the dynamics given Eq. \ref{eq:phonons}, in the presence of a phenomenological damping $\gamma$. Near resonance, Eq. \ref{eq:phonons} is simplified to a driven dissipative Kerr oscillator, a canonical model in quantum non-linear optics \cite{drummond, milburn, ciuti} and nano-mechanics \cite{dykman}. For resonant driving and strong fields $E_0 > E_{\rm th} = \sqrt{\frac{M}{\beta}} \gamma$, the driven phonon---starting from a small inversion symmetry breaking initial displacement arising from the Raman term---becomes parametrically unstable, growing exponentially  in oscillation amplitude until anharmonic non-linearities stabilize it \cite{milburn}. 

At steady state, the phonon \textit{coherently} oscillates with $Q(t) = Q_{\rm ss} \cos(\omega_{\rm dr}t)$.  A semi-classical analysis gives the steady-state amplitude $Q_{\rm ss} = \sqrt{\frac{1}{\lambda}\big(\sqrt{\kappa^2-\gamma^2} -\Delta \big)}\ell_0$\cite{milburn}; where $\kappa = \beta E_0^2/M\omega_{\rm eff}$; where the detuning $\Delta = \omega_{\rm eff}-\omega_{\rm dr}$; where $\omega_{\rm eff} = \sqrt{\omega^2+\beta E_0^2/M}$, capturing the shift in the phonon frequency arising from the DC component of the parametric drive. We assume, hereafter $\beta>0$ (the phonon hardens \footnote{For $\beta<0$, the phonon softens, increasing the bipolaron energy $U_{\rm bi} = g^2/(2 M \omega_{\rm eff}^2)$, a more trivial flavor of photo-enhanced pairing.}). A linear stability analysis within the semi-classical approximation supports the classical expectation that such a steady state is (meta)stable so long as  $|\Delta| <\sqrt{\kappa^2-\gamma^2}$. 

We now attend, briefly, to two passing remarks. Inversion symmetry dictates that if $Q_{\rm ss} \cos(\omega_{\rm dr}t)$ is a steady state, so must  $-Q_{\rm ss} \cos(\omega_{\rm dr}t)$ be. Ergodicity is recovered from the two underlying, phase-shifted, metastable steady-states via quantum-noise activated switching\cite{dykman}. However, for sufficiently large $E_0$---thereby  $Q_{\rm ss}$---over-barrier activation is exponentially suppressed\cite{dykman} and therefore, over the course of the experiment, one can safely consider a single (meta)stable steady state. Moreover, while at the onset of parametric amplification ($E_0 \sim E_{\rm th}$) the steady-state dynamics are meaningfully squeezed\footnote{This can be understood pictorially. Before the instability, the Wigner function in the rotating frame of the drive is a symmetric Gaussian; afterward it becomes a bimodal Gaussian with peaks at $\pm Q_{\rm ss}$. Only in the shearing process to achieve this is substantial squeezing generated.}, for fields such that $Q_{\rm ss} \gg \ell_0$,  $g^2(0) \approx 1$ \cite{ciuti}: The phononic state is essentially a coherent state. 

We now specify $H_{\rm e-ph}$. As oscillations of the phonon displacement get larger, they start exploring even weak non-linearities present in the electron-phonon coupling. We minimally model this by examining the \textit{non-linear Holstein model}:
\begin{equation}
    H_{\textrm{e-ph}} = g \sum_{i} Q_{i}\big(1+ \epsilon  Q_i \ell_{0}^{-1} \big)n_i,
\label{eq:non_lin_holstein}
\end{equation}
where $\epsilon$ quantifies the relative strength between the linear and quadratic electron-phonon couplings and $n_i$ is deviation around the average filling, i.e. $\langle n_i \rangle =0$. Previous \textit{ab initio} estimates of $\epsilon$---albeit in a different material context---lie between $0.01-0.2$\cite{nonlin-e-ph}. While Holstein coupling is not the correct form of the electron-phonon coupling for the Jahn-Teller $H_g$ modes, we believe distinction adds obfuscating rococoism\footnote{In alkali-doped fullerides, the relevant optical Raman modes are the 8 $H_g$ Jahn-Teller phonons which, for each mode, couple with five different polarizations across the three orbitals. We elide this distinction for a few reasons. Primarily, we would like to present our insights without any superfluous complexity, arising from, for example, having to deal with mutli-band superconductivity---extensions of our ideas to mutli-band interorbital electron-phonon systems are straightforward but tedious. Second, the most relevant distinctions between Holstein and Jahn-Teller coupling for superconductivity arise from two subtleties: (i) electronic Berry phases which enhance the bipolaron binding energy only at \textit{weak} coupling\cite{assa}; (ii) the subtle interplay between electron-phonon superconductivity and strong, unretarded Coulomb interactions---while Jahn-Teller mediated pairing in enhanced near the Mott transition due to bandwidth suppression, Holstein mediated does not survive the interaction induced suppression of charge fluctuations\cite{hanK3C60, Gunnarsson_04}. For our purposes, the first point is not so relevant as the linear electron-phonon coupling is strong; the second point is beyond the scope of our analysis as we do not consider the effects of Coulomb interactions---one can instead imagine that Eq. \ref{eq:BasicModelFull} is a minimal description of electron-like quasi-particles after they are dressed by the strong Coulomb interactions.}, obscuring the essential physics by adding superfluous complexity. 

Building on our understanding of the steady-state dynamics of the driven Raman coordinate, we linearize the fluctuations of $Q$ around its instantaneous expectation value $\langle Q(t) \rangle$ as: $Q = \langle Q(t) \rangle + \tilde{Q}$. Expanding the electron-phonon interaction to lowest non-trivial order in $\tilde{Q}$, we obtain:
\begin{equation}
    H_{\textrm{e-ph}}(t) = g (1+A(t)) \sum_{i} \tilde{Q}_{i}n_i,
\label{eq:time_dependent_holstein}
\end{equation}
where $A(t) = 2\epsilon Q_{\rm ss} \ell_0^{-1} \cos(\omega_{\rm dr} t) \equiv A_0  \cos(\omega_{\rm dr} t)$ --- a \textit{time-dependent linear} Holstein model. 

Having derived Eq.~\ref{eq:time_dependent_holstein} from microscopic considerations, we digress from the \textit{full} microscopic setting and unpack the implications that a time-dependent effective electric phonon coupling has on non-equilibrium superconductivity. After building intuition for this time-dependent Holstein model, we will revert our gaze to the microscopic setting. Accordingly, for the moment, we ignore the renormalization of the phonon frequency due to the drive and consider the physics arising from a fixed  $A_0$, independent of $\omega_{\rm dr}/\omega$, ignoring its origin from an underlying parametric resonance. 

Our project will be to derive the effective electron-electron interaction arising from phonon-fluctuations using a time-dependent Schrieffer-Wolff Transformation.  Let us briefly review such a transformation. Given a time-dependent $H(t) = H_0 + V(t)$, where $V(t)$ is a time-dependent perturbation (e.g. $H_{\rm e-ph}(t)$), we would like to move to a rotating frame $U(t) = \exp(i S(t))$, where the lowest order interaction between different systems (e.g., electrons and phonons) is ``integrated out''. Specifically, this means choosing $S(t)$ such that $V(t) + i[H_0, S(t)] + \partial_t S(t) =0$. The effective interaction generated by this procedure is given by $\frac{i}{2}[V(t), S(t)]$. 

As we are working in the anti-adiabatic limit (i.e. expanding around $\omega \to \infty$), we wish to eliminate terms to lowest non-trivial order in $\omega$. This can be done by neglecting $H_{\rm el}$ and taking $H_0 = \tilde{H}_{\rm ph} = \frac{P^2}{2 M }+ \frac{M \omega^2}{2} \tilde{Q}^2$ and $V(t) = H_{\rm e-ph}(t)$, i.e. solving the local problem. The local approximation is in keeping with the standard treatment of the electron-phonon problem in \ch{K3C60}, which introduces a local anti-Hunds coupling arising from integrating out high-frequency molecular Jahn-Teller fluctuations\cite{assa, nomura}.

%\footnote{By making the local approximation, we are neglecting  polaronic renormalization which---on the weak coupling (BCS) side---suppresses tunneling, enhances the density of states, and thereby is typically favorable for pairing \cite{knap_dynamical_2016, babadi_theory_2017}. As this effect is both well characterized and further supports high-temperature pairing, we neglect it in our treatment.}.

Within the local approximation, the lowest order terms can be removed by choosing $S(t) = \big(\lambda_P(t) P_i + \lambda_Q(t) \tilde{Q}_i\big) n_i$ where $\lambda_P(t) = \frac{g}{M \omega^2}\big(1+ A_0 \mathcal{R}(\omega_{\rm dr})\big)$ and  $\lambda_Q(t) = - M \frac{d}{dt}\lambda_P(t)$, where $\mathcal{R}(\omega_{\rm dr}) = \frac{\omega^2}{\omega^2-\omega_{\rm dr}^2}$. The time dependent electron-electron pairing interaction that arises from this is given  by:
\begin{equation}
U(t) = \frac{g^2}{2 M\omega^2}\big(1+ A_0 \cos(\omega_{\rm dr}t)\big)\big(1+A_0 \cos(\omega_{\rm dr}t) \mathcal{R}(\omega_{\rm dr})\big).
\label{eff_attraction}
\end{equation}

We underscore the presence of $\mathcal{R(\omega_{\rm dr}})$ in Eq.~\ref{eff_attraction} resonantly enhances $U(t)$ \textit{in addition to} any enhancement of $A_0$ that may arise from the underlying parametric resonance, enabling a \textit{giant} resonant enhancement of the superconducting photo-susceptibility. To avoid divergences for $\omega_{\rm dr} \sim \omega$, we perform our analysis in the presence of an Ohmic bath which gives rise to a finite phonon dissipation rate $\gamma$. In the presence of dissipation, $\mathcal{R(\omega_{\rm dr})} \to \omega^2 \big(\big(\omega^2-\omega_{\rm dr}^2\big)^2+\gamma^2 \omega_{\rm dr}^2\big)^{-1/2}$ with a corresponding phase lag $\tan (\theta(\omega_{\rm dr}))= \gamma \omega_{\rm dr}(\omega^2-\omega_{\rm dr}^2)^{-1}$ such that the term $A_0 \cos(\omega_{\rm dr}t)R(\omega_{\rm dr}) \to A_0 \cos(\omega_{\rm dr}t+\theta(\omega_{\rm dr}))R(\omega_{\rm dr})$ (see Appendix B for a systematic  derivation). On resonance,  $R(\omega_{\rm dr} =\omega) = \mathcal{Q}\equiv \omega / \gamma$. Sufficiently strong \textit{modulations} of the attractive interaction will trigger a Floquet Cooper pairing instability at electronic temperatures $T_{c}^\star$ much larger than $T_c$ \cite{knap_dynamical_2016, babadi_theory_2017}. 

\subsection{Appendix B: Dissipative Schrieffer-Wolff in an interacting electron-phonon system}

We begin with a brief review of the canonical treatment of a local boson interacting with a bosonic bath (i.e., the Caldeira-Leggett problem), to facilitate the discussion of the dissipative Schrieffer-Wolff used to tackle the local driven electron-phonon problem in the presence of a bosonic bath. 

We start with the Hamiltonian: 
\begin{equation}
   H_0 = \sum_i \frac{M \omega^2}{2} Q_i^2 + \frac{P_i^2}{2 M } + \sum_{i,l}\lambda_{l} Q_i X_{i,l} + \sum_{l} \left( \frac{\omega^2_l}{2} X_{i,l}^2 + \frac{1}{2} \Pi_{i,l}^2 \right),
   \label{eq:H0}
\end{equation}

where $Q_i$ is a dispersion-less phonon mode per lattice site, $i$ and $X_{i,l}$ is a harmonic bath where $l$ corresponds to the bath harmonic oscillators at position $i$. The different operators, obey canonical commutation relations, $\left[Q_i, P_j \right] = i \delta_{i,j}$ and $\left[ X_{i,l}, \Pi_{j,l'}\right] = i \delta_{i,j}\delta_{l,l'}$. This linear model can be used to simulate disspation damping of phonon modes within a Hamiltonian treatment. Classical equations of motion are given by:
\begin{align}
    \partial_t \langle Q_i \rangle =& \frac{\langle P_i \rangle }{M}, \label{eq:Expe1}\\
    \partial_t \langle P_i \rangle =& - M \omega^2 \langle Q_i \rangle - \lambda_{l} \langle X_{i,l} \rangle ,\label{eq:Expe2}\\
    \partial_t \langle X_{i,l} \rangle =& \langle P_{i,l} \rangle, \label{eq:bath1}\\
    \partial_t \langle \Pi_{i,l} \rangle =& -  \omega^2_{l} \langle X_{i,l} \rangle - \lambda_{l} \langle X_{i,l} \rangle . \label{eq:bath2}
\end{align}

Eliminating the bath degrees of freedom by using equations~(\ref{eq:bath1}) and (\ref{eq:bath2}) to replace $\langle X_{i,l} (t)  \rangle = \lambda_{l} \int d t' C_l (t - t') \langle Q_i(t') \rangle $. The final equation of motion for $\langle Q_i \rangle$ is given by:
\begin{equation}
    \partial_t^2\langle Q_i \rangle(t) + \omega^2 \langle Q_i \rangle (t) + \sum_l \frac{\lambda_{l}^2}{M} C_l(t-t') \langle Q_i \rangle(t') = 0.
\end{equation}
The Fourier transform of $C_l(t-t')$ is given by: 
\begin{equation}
    C_l(\Omega) = \frac{1}{ \Omega^2 + i \eta \Omega - \omega_{l}^2}. 
\end{equation}
In the limit of a continuum of modes inside the bath, the sum over $l$ can be replaced by an integral of the type: 
\begin{equation}
    f(\Omega) = \int d\epsilon \frac{\lambda^2(\epsilon)}{M} \frac{1}{\Omega^2 + i \eta \Omega - \epsilon^2} \rho(\epsilon),
\end{equation}
where $\rho(\epsilon)$ is the density of states of the continuum. Assuming the $\lambda^2(\epsilon)\rho(\epsilon) $ decay fast enough as $|\epsilon| \rightarrow \infty $ and has no poles, we can perform the contour integral to evaluate $f(\Omega)$ :

\begin{equation}
    f(\Omega) = - \frac{2 \pi i\lambda^2(\Omega) \rho(\Omega)}{2 M \Omega} + \frac{2 \pi i\lambda^2(-\Omega) \rho(- \Omega)}{- 2 M \Omega} = i \frac{\pi \left(\lambda^2(\Omega) \rho(\Omega) + \lambda^2(-\Omega) \rho(-\Omega) \right) }{M \Omega}.
\end{equation}

Finally, assuming that $\rho(\Omega) \lambda^2(\Omega) $ is a real and even function of $\Omega$ we get that the function, $f(\Omega)$ is both odd in frequency, $f(- \Omega) = - f(\Omega) $ and purely imaginary, $f(\Omega)^* = - f(\Omega)$. As a result $f(\omega)$ acts as a generalized dissipative force. The simplest example, is assuming $\frac{\pi \left(\lambda^2(\Omega) \rho(\Omega) + \lambda^2(-\Omega) \rho(-\Omega) \right) }{M \Omega} = 2\pi \gamma \Omega  $, which is ohmic dissipation. In time this can be represented as: \begin{equation}
    \partial_t^2\langle Q_i \rangle(t) + \gamma\partial_t  \langle Q_i \rangle (t)+ \omega^2 \langle Q_i \rangle (t) = 0,
\end{equation}
the prototypical example of a damped harmonic oscillator with dissipation quantified by $\gamma$. 

Leveraging the same set-up as in the previous section, we now wish to perturbatively diagonalize the electron-phonon interacting Hamiltonian while including the linear coupling to the bath. We begin by expositing the \textit{time-dependent} Holstein model in which the interacting Hamiltonian is given by:
\begin{equation}
    H_{\rm int} = g(t) \sum_i c_i^\dag c_i Q_i ,
\end{equation}
where in the local limit hopping between sites is neglected, as discussed in the main text. For the purposes of this section, we suppress the distinction between the phonon coordinate and its fluctuation and write $Q$ and $\tilde{Q}$. 

Using time-dependent Schrieffer-Wolff approach, the effective Hamiltonian in a rotating frame given by $U = \mbox{exp} (i S) $ is given by:

\begin{equation}
    H_{\rm eff} = U^\dag H U - i U^\dag \partial_t U = H_0 + H_{\rm int} + i [H_0,S] + \partial_t S + i [H_{\rm int}, S] - \frac{1}{2} [ [ H_0,S] - i \partial_t S, S]   
\end{equation}

where $H_0$ is defined in equation~(\ref{eq:H0}). For treating the time-derivative of $U$, we use the formula:
\begin{equation}
   \partial_t U =  \partial_t e^{ i S} = e^{i S} \int_0^1 d u e^{- i u S } i \partial_t S e^{ i u S} = U \int_0^1 du\left( i \partial_t S - [\partial_t S, S] u + \dots \right) = U \left( i \partial_t S - \frac{1}{2} [\partial_t S, S]\right) .
\end{equation}

To diagonalize the Hamiltonian to first order in the electron phonon coupling, $g$, we require:
\begin{equation}
    i [H_0, S] +  \partial_t S = - H_{\rm int}. \label{eq:SWcondition}
\end{equation}
Using this relationship, the effective Hamiltonian to second order in $g$ is given by:
\begin{equation}
    H_{\rm eff} = H_0 + \frac{i}{2}[ H_{\rm int}, S].
\end{equation}
To satisfy equation~(\ref{eq:SWcondition}), we choose $S$ to be:

\begin{equation}
    S = \alpha(t) \sum_i Q_i c_i^\dag c_i - \beta(t) \sum_i P_i c_i^\dag c_i  + \sum_l \gamma_l(t) \sum_i X_{i,l} c_i^\dag c_i - \sum_l \delta_l(t) \sum_i \Pi_{i,l} c_i^\dag c_i 
\end{equation}

Using the above transformation, equation~(\ref{eq:SWcondition}), becomes:
\begin{align}
    M \omega^2 \beta(t) + \sum_l \lambda_l \delta_l 
    + \partial_t \alpha(t) + g(t) =& 0, \\
    \frac{1}{M} \alpha(t) - \partial_t \beta(t) =& 0,\\
    \omega_l^2 \delta_l(t) + \lambda_l \beta (t) + \partial_t \gamma_l(t) =& 0, \\
    \gamma_l(t) - \partial_t \delta_t =&0,
\end{align}
Once can confirm that the above equations are the same as equations (\ref{eq:Expe1})-(\ref{eq:bath2}) with the identification of $\expe{Q_i} \rightarrow \beta(t)$, $\expe{P_i} \rightarrow \alpha(t)$, $\expe{X_{i,l}} \rightarrow \delta_l(t)$ and $\expe{\Pi_{i,l}} \rightarrow \gamma_l(t)$ and in the presence of a driving term given by $g(t)$. Using the same definitions as in the previous section, $\beta(t)$ is found to be:
\begin{equation}
    \partial_t^2 \beta(t)+ \sum_l \frac{\lambda_l^2}{M} \int dt' C_l(t - t' ) \beta(t') + \omega^2 \beta(t) = - \frac{g(t)}{M} . 
\end{equation}
Assuming an ohmic bath we identify $\sum_l \frac{\lambda_l^2}{M} \int dt' C_l(t - t' ) \beta(t') = \gamma \partial_t \beta(t)$ to arrive at the equation:
\begin{equation}
    \partial_t^2 \beta(t)+ \gamma \partial_t \beta(t) + \omega^2 \beta(t) = - \frac{g(t)}{M}.
\end{equation}
Knowing $\beta(t)$ allows to determine $H_{\rm eff}$ which is given by:
\begin{equation}
    H_{\rm eff} = H_0 - \frac{1}{2} g(t)\int dt' D(t - t') g(t') \sum_{i,\sigma,\sigma'} c^\dag_{i,\sigma} c_{i,\sigma} c^\dag_{i,\sigma'} c_{i,\sigma'} ,
\end{equation}
where $D(t -t')$ is the classical response function of the coordinate $X_{i}$, given by the equation:
\begin{equation}
    \left( \partial_t^2 + \gamma \partial_t + \omega^2 \right) D(t - t') = \frac{1}{M} \delta(t - t')
\end{equation}

\end{document}